\documentclass[twocolumn]{aastex62}

\usepackage{epsf}
\usepackage{amsmath}                
\usepackage{amsfonts}               
\usepackage{amssymb}                
\usepackage{epsfig}                 
\usepackage{float}
\usepackage{color}
\usepackage{tabularx}
\usepackage{comment}
\usepackage{makecell}
\usepackage{gensymb}               

\def \km{~\rm{km}}

\def \erg{~\rm{erg}}

\def \yr{~\rm{yr}}

\begin{document}

\title{Simulating the negative jet feedback mechanism in common envelope jets supernovae}

\author{Aldana Grichener}
\affiliation{Department of Physics, Technion, Haifa, 3200003, Israel; aldanag@campus.technion.ac.il; soker@physics.technion.ac.il}

\author{Coral Cohen}
\affiliation{Department of Education in Science and Technology, Technion, Haifa, 3200003, Israel; coralcohen@campus.technion.ac.il}

\author[0000-0003-0375-8987]{Noam Soker}

\affiliation{Department of Physics, Technion, Haifa, 3200003, Israel; aldanag@campus.technion.ac.il; soker@physics.technion.ac.il}
\affiliation{Guangdong Technion Israel Institute of Technology, Shantou 515069, Guangdong Province, China}

\begin{abstract}
We use the stellar evolution code \textsc{MESA} to study the negative jet feedback mechanism in common envelope jets supernovae (CEJSNe) where a neutron star (NS) launches jets in the envelope of a red supergiant (RSG), and find that the feedback reduces the mass accretion rate to be $\chi_{\rm j} \simeq 0.04-0.3$ times the mass accretion rate without the operation of jets. We mimic the effect of the jets on the RSG envelope by depositing the energy that the jets carry into the envelope zones outside the NS orbit. The energy deposition inflates the envelope, therefore reducing the density in the NS vicinity, which in turn reduces the mass accretion rate in a negative feedback cycle. 
In calculating the above values for the negative jet feedback coefficient (the further reduction in the accretion rate) $\chi_{\rm j}$, we adopt the canonical ratio of jets power to actual accretion power of $0.1$, and the results of numerical simulations that show the actual mass accretion rate to be a fraction of $0.1-0.5$ of the Bondi-Hoyle-Lyttleton mass accretion rate.   
\end{abstract}

\keywords{stars: jets -- stars: massive -- stars: neutron -- binaries: close -- stars: supergiants}

\section{Introduction}
\label{sec:intro}

Common envelope jets supernovae (CEJSNe) are transient events in which a neutron star (NS) or a black hole (BH) launch jets as they spiral in inside the envelope of a red supergiant (RSG) star and later thorough its core \citep{SokerGilkis2018} . When inside the RSG, the compact object accretes mass from its surroundings through an accretion disk. The accretion disk launches two jets that propagate more or less perpendicular to the orbital plane in opposite directions. The jets deposit their kinetic energy in the giant star and eventually explode it in a CEJSN event. If the compact object does not penetrate the core then the transient event is termed CEJSN impostor.

A CEJSN event resembles a core collapse supernova (CCSN) in many respects. The explosion energy in both cases is $\approx 10^{50}-10^{52} \erg$ and the ejecta mass about equals the RSG envelope mass. However, the spiraling-in process of the NS/BH in the RSG envelope ejects some mass with which the ejecta collide later. Therefore, the light curve itself resembles those of energetic CCSNe with circumstellar matter (CSM).  

The scenario that leads to a CEJSN (see for instance Fig. 1 in \citealt{GrichenerSoker2019a}) begins with two massive main sequence stars, one of them heavier than the other. The more massive star $M_1$ evolves into a RSG while its companion $M_2$ is still on the main sequence. The remaining main sequence star is massive enough to spin-up the envelope of the RSG to be synchronized with the orbital motion such that the tidal force vanishes. This prevents further spiraling-in, i.e. prevents a common envelope evolution (CEE) at this stage. The RSG keeps evolving until it explodes as a CCSNe, leaving a NS/BH behind. If the natal kick of the compact object (mainly the NS) at this stage is not too high, and if enough mass remains bound to the binary system, the CCSN explosion leaves a binary system of a NS/BH and a main sequence star. The main sequence star eventually evolves into a second RSG that can engulf the NS/BH initiating a CEE stage that might lead to a CEJSN event. The initially  more massive star should be of mass $M_1 \ga 8.5M_{\rm \odot}$ to form a NS, while its companion should also be massive since the NS is required to spiral-in into the envelope of $M_2$ as it becomes a RSG, and for that it cannot force the RSG envelope to synchronization. We estimate the condition to be that its initial mass is $M_2 \ga 6 M_\odot$ as it can grow by mass accretion from the primary before it evolves to become a RSG. However, to prevent fine tuning with the other parameters, like initial semi-major axis and eccentricity, we consider CEJSNe to results from more massive stars, i.e., initial masses of $M_1 \ga 15 M_\odot$ and $M_2 \ga 10 M_\odot$. 

Numerous studies of CEJSNe and CEJSN impostors were conducted over the past few years (e.g., \citealt{SokerGilkis2018};  \citealt{Gilkiselal2019}; \citealt{Sokeretal2019}; \citealt{GrichenerSoker2019a}; \citealt{LopezCamaraetal2019}; \citealt{LopezCamaraetal2020MN} \citealt{GrichenerSoker2021}; \citealt{Soker2021}). Other studies explore the properties of NS/BH - core mergers, yet do not explicitly include jets (e.g., \citealt{FryerWoosley1998}; \citealt{Chevalier2012}; \citealt{Schroderetal2020}). 
The most generic property of CEJSNe is non-monotonic light curves. This results from the early  non-spherical ejection of envelope gas that forms a relatively dense CSM and from the operation of jets that might have variable intensity. One natural outcome of the envelope mass ejection by jets is that the CEE efficiency parameter can become larger than unity. Indeed, some scenarios of NS/BH spiraling-in inside RSG envelopes require values of $\alpha_{\rm CE} > 1$ (e.g. \citealt{Fragosetal2019, BroekgaardenBerger2021, Zevinetal2021, Garciaetal2021}). We note that there is an ongoing debate on the accretion rate by the NS/BH relative to the Bondi-Hoyle-Lyttleton mass accretion rate (see for instance \citealt{Schreieretal2021} for a discussion). More research is needed to better constrain all parameters of CEJSNe. 

Due to their wide diversity CEJSNe might be responsible for various astrophysical phenomena from rare peculiar explosions such as iPTF14hls \citep{SokerGilkis2018} and AT2018cow \citep{Sokeretal2019}, to more general processes in high energy astrophysics, such as heavy r-process nucleosynthesis in the early universe (\citealt{GrichenerSoker2019a}; \citealt{GrichenerSoker2019b}) and the PeV neutrinos emission detected by IceCube in 2013 \citep{GrichenerSoker2021}. We note that many of the scenarios for double NSs (e.g., \citealt{VignaGomezeal2018}), double BHs (e.g., \citealt{Mapelli2020}) and NS-BH binary (e.g., \citealt{Hoangetal2020}) formation and merger involve a phase where a NS/BH spirals in inside a RSG envelope. Therefore, many of them must evolve through a CEJSN impostor phase. This further stresses the increasing importance of CEJSNe and CEJSN impostors and the ongoing research in the field.

In the present study we use approximate methods to estimate the effect of envelope inflation on the accretion rate when a NS launches jets in the envelope of a very massive RSG. The jets that a compact companion launches inside the envelope of a giant star operate in a feedback cycle that has both a positive and a negative component (see \citealt{Soker2016Rev} for a review). The positive feedback comes from the removal of energy and angular momentum from the immediate vicinity of the accreting compact star. Since jets are likely to remove high entropy gas at large velocities (about the escape speed from the compact object and more), they reduce the pressure near the accreting object and allow more gas to flow-in, increasing the mass accretion rate (e.g., \citealt{Shiberetal2016, Chamandyetal2018}). In the case of CEJSNe where the accreting object is a NS, neutrino cooling carries most of the energy. This is the positive feedback component of the cycle in the sense that a low accretion rate does not allow for neutrino cooling to occur \citep{HouckChevalier1991, Chevalier1993, Chevalier2012}. 

The negative component of the feedback cycle comes from the energy and momentum that the jets deposit to the common envelope matter (CEM) as the compact star orbits inside the envelope (\citealt{Soker2016Rev}). This removes CEM from the vicinity of the compact object by both direct local effects and by the inflation of the entire envelope. The decrease in the local density reduces the accretion rate, closing the negative feedback cycle. 

In section \ref{sec:Mimiking} we describe our assumptions and approximated methods to mimic the negative feedback cycle,  and portray the numerical simulation of energy injection into a spherical RSG envelope. In section \ref{sec:Results} we describe our results. We summarise our main conclusions in section \ref{sec:Summary}. 

\section{Mimicking the negative feedback mechanism}
\label{sec:Mimiking}
 
We examine the approximate influence of the jet feedback mechanism on the mass accretion rate and jets power by making several approximations and assumptions, as we list below. We perform one-dimensional simulations of a RSG to find the mass accretion rate onto a NS in a CEJSN event. However, to truly explore the jet feedback mechanism (section \ref{sec:intro}) one needs to conduct full three-dimensional hydrodynamical simulations that include jets in a CEE (e.g., \citealt{Hilleletal2021}). 

\subsection{Mass accretion rate}
\label{subsec:AccretionRate}

We take the actual NS mass accretion rate to be a fraction $\xi$ of the Bondi-Hoyle-Lyttleton mass accretion rate (\citealt{HoyleLyttleton1939}; \citealt{BondiHoyle1944})
\begin{equation}
\dot M_{\rm BHL} = \pi R^2_{\rm acc} v_{\rm rel} \rho, 
\label{eq:BHLaccretion}
\end{equation} 
where $R_{\rm acc}$ is the accretion radius, $v_{\rm rel}=v_{\rm K}-v_{\rm rot}$ is the NS-envelope relative velocity, $v_{\rm K}$ is the Keplerian velocity of the NS inside the RSG, $v_{\rm rot}$ is the envelope rotation velocity and  $\rho$ is the envelope density at the location of the NS. We neglect the rotation of the envelope such that $v_{\rm rel} \simeq v_{\rm K}$.  This is a plausible assumption since the NS cannot bring the envelope to synchronization before it enters the envelope. Namely, already when the NS is on the RSG surface, $a=R_{\rm RSG}$, we have the inequality $v_{\rm rot} (R_{\rm RSG})< v_{\rm Kep}(R_{\rm RSG})$. We assume a solid body rotation of the envelope $\Omega_{\rm env} ={\rm constant}$, such that $v_{\rm rot} = \Omega_{\rm env} r$. The Keplerian velocity varies as $\propto r^{-1/2}$. At the location of the NS $r=a$ we find that $v_{\rm rot}(a)/ v_{\rm Kep}(a) \simeq  (a/R_{\rm RSG})^{3/2} [v_{\rm rot}(R_{\rm RSG})/ v_{\rm Kep}(R_{\rm RSG})] < (a/R_{\rm RSG})^{3/2}$. Therefore, when the NS is deep inside the envelope such that $a \la 0.5 R_{\rm RSG}$ we can neglect the rotation of the envelope. Moreover, to the accuracy of our study the expression for $R_{\rm acc}$ takes the form
\begin{equation}
R_{\rm acc} \simeq 
\frac {2 G M_{\rm NS}} {v^2_{\rm K}}. 
\label{eq:BHLradius}
\end{equation} 

Three-dimensional hydrodynamical simulations over the last 4 decades (e.g., 
\citealt{Livioetal1986}; \citealt{RickerTaam2008}; \citealt{Chamandyetal2018})
have shown that the actual accretion rate onto a NS is $\dot M_{\rm acc} =  \xi \dot M_{\rm BHL} $ where in most cases $\xi \simeq 0.1-0.5$, yet smaller values are also possible (e.g., \citealt{MacLeodRamirezRuiz2015a}; \citealt{MacLeodRamirezRuiz2015b}). 
 \citealt{Chamandyetal2018} consider that the jets inflate low density cocoons that in turn lead to a lower accretion rate by the mass-accreting body in a negative feedback cycle. The other studies do not take into account the effect of the jet feedback mechanism.

In the present study we consider the role of jets in reducing the accretion rate by
inflating the RSG envelope, hence reducing the density at the vicinity of the NS by a factor of
\begin{equation}
\chi_{\rm j} \equiv \frac {\rho}{\rho_{\rm 0}} ,
\label{eq:ChiJ}
\end{equation}
where $\rho$ is the density at the location of the NS in the inflated envelope, and $\rho_0$ is the density in the same location in the unperturbed envelope. This implies that the actual accretion rate due to the effect of the jets would be 
\begin{equation}
\dot M_{\rm acc,j} = \chi_{\rm j} \dot M_{\rm acc,0}=  \chi_{\rm j} \xi \dot M_{\rm BHL,0},  
\label{eq:AccretionRate}
\end{equation}
where the subscript `0' indicates values in the unperturbed envelope. 
The correct radius to evaluate the densities in equation (\ref{eq:ChiJ}) is at the location of the NS. However, in the spherical model that we use in this study, we cannot follow the spiraling-in orbit of the NS and the propagation of the jest inside the envelope since the location of the NS and the jets introduce high-non-spherical flows that our spherical model cannot account for (for these non-spherical effects see \citealt{Hilleletal2021}). Therefore, We will consider a general evolution of several years during which the NS plunges deep into the envelope, and examine the value of $\chi_{\rm j}$ at three different radii deep inside the envelope. 

\subsection{The power of the jets}
\label{subsec:JetsPower}

The energy that the two jets, one on each side of the equatorial plane, deposit to the RSG envelope is 
\begin{equation}
\dot E_{\rm 2j} = \eta \frac {G M_{\rm NS} \dot M_{\rm acc,j}}{R_{\rm NS}} = 
 \zeta \frac {G M_{\rm NS} \dot M_{\rm BHL,0}}{R_{\rm NS}}  , 
\label{eq:JetsPower}
\end{equation}
where $M_{\rm NS}=1.4M_{\rm \odot}$ and $R_{\rm NS}=12 \km$ are the mass and the radius of the NS used in this study, respectively, and 
\begin{equation}
\zeta \equiv \eta \chi_{\rm j} \xi.
\label{eq:Zeta}
\end{equation}
We take the efficiency parameter to be $\eta = 0.1$ (e.g., \citealt{Schroderetal2020}), and perform several simulations in which we inject jets energies with different values of $\zeta$ to the RSG envelope (subsection \ref{subsec:Numerics}). In each simulation we fix the value of $\zeta$, and we determine the value of $\chi_{\rm j}$ by equation (\ref{eq:ChiJ}) at three orbital separations after energy deposition to the envelope. We then compute the value of $\xi$ from $\xi = {\zeta}/{\eta \chi_{\rm j}}$ (section \ref{sec:Results}).

\subsection{The Numerical scheme and model}
\label{subsec:Numerics}

We use the stellar evolution code \textsc{MESA} (e.g., \citealt{Paxtonetal2010}; \citealt{Paxtonetal2013}; \citealt{Paxtonetal2015}; \citealt{Paxtonetal2018}; \citealt{Paxtonetal2019}) to obtain a non-rotating spherical massive RSG model with zero age main sequence (ZAMS) mass of $M_{\rm ZAMS}=60M_{\rm \odot}$ and a metalicity of $Z=0.02$. We let the star evolve until it reaches a radius of $R_{\rm RSG}=800R_{\rm \odot}$. At this stage, the stellar mass is $M_{\rm RSG}=59.8M_{\rm \odot}$ due to mass loss by stellar winds. We assume that the RSG swallows a NS of mass $M_{\rm NS}=1.4M_{\rm \odot}$ at this point. The NS launches jets inside the envelope of the RSG, driving it to expand due to deposition of the kinetic energy of the jets to the envelope as they collide with the envelope gas. We assume that the NS spirals-in throughout the envelope in a constant rate until the end of our simulations. Hence, we inject the jets energy into the envelope in spherical shells and with a constant power per unit mass over a time of $t=0$ to $t=\tau_{\rm j}$. 

At each time step $\Delta t$ we inject the energy of the jets $\Delta E_{\rm 2j} = \dot E_{\rm 2j} \Delta t$ into a thick spherical shell bounded from inside by the sphere $r_{\rm in}=a(t)$, i.e., the location of the NS, and from outside by the sphere $r_{\rm out}=700R_{\rm \odot}$, i.e., a sphere somewhat below the surface of the RSG before the onset of the CEE. Due to envelope expansion the density decreases with time, and we calculate at each time step the mass $M_{\rm TS}$ in a thick shell into which we deposit the energy of the jets. We assume that the jets propagate to large distances, and therefore distribute the energy in a thick spherical shell from the location of the NS at each time step to the above value of $r_{\rm out}$.
Into each spherical numerical shell of mass $\Delta M_{\rm shell}$ we inject during this time step an energy of $\Delta E_{\rm shell}= (\Delta M_{\rm shell}/M_{\rm TS}) \Delta E_{\rm 2j}$. Typically there are hundreds up to about a thousand numerical shells into which we deposit the energy.

We perform two sets of simulations. In the first set, we inject energy to the envelope using equation (\ref{eq:JetsPower}), until the NS reaches an orbital separation of $a=100R_{\rm \odot}$. We assume that the time it takes the NS to spiral in to that radius is $\tau_{\rm j}=5\yr$, which is $5.4$ times the Keplerian orbital period on the surface of the RSG. Hence, we inject the energy from the location of the NS to $r_{\rm out}=700R_{\rm \odot}$ during this time. We conduct ten different simulations for ten equally spaced values of $\zeta$ in the range $\zeta = 5\times10^{-4} - 5\times10^{-3}$. In the second set of simulations we let the NS dive deeper, and inject the energy until the NS reaches an orbital separation of $a=20R_{\rm \odot}$. As the spiral-in slows down at deeper layers of the envelope (e.g., \citealt{Passyetal2012, RickerTaam2012, Ivanovaetal2013, Sandetal2020, GlanzPerets2021}), we take the energy injection time to be $\tau_{\rm j}=10\yr$ in this case. We conduct ten simulations with the same values of $\zeta$ as in the previous set.  

\subsection{Neglecting the orbital energy}
\label{subsec:NeglectingOrbital}

To reveal the role of the jets we neglect the effect of the orbital energy that the core of the RSG star and the NS release as they spiral-in towards one another in a CEE, $E_{\rm orb}(r) \simeq GM_{\rm RSG}(r) M_{\rm NS}/2r$. We use a massive RSG model such that $E_{\rm orb}(r)$ in the orbital separations range where we follow the spiraling-in NS is not only smaller than the energy that the jets deposit in the RSG envelope, $E_{\rm 2j}(r)$, but it is also smaller than the binding energy of the envelope in this volume. This assures us that the orbital energy does not change much the envelope, which is inflated mainly by the energy that the two jets deposit as they collide with the envelope gas.    

The energy that the jets deposit to the RSG envelope until the NS reaches the  radii $r=20R_{\rm \odot}$ and $r = 100 R_\odot$ are (for our different models) $E_{\rm 2j}(20R_{\rm \odot}) = 5.5\times 10^{50} - 5.5\times 10^{51} \erg$ and $E_{\rm 2j}(100R_{\rm \odot}) = 5.7\times 10^{49} - 5.7\times 10^{50} \erg$, respectively. Although we take the duration of the spiraling-in to $r=20R_\odot$ to be twice as long as that to $r=100 R_\odot$, the extra spiraling-in takes place in a much denser envelope that results in a much larger mass accretion rate and hence in jets with much higher powers. Therefore $E_{\rm 2j}(20R_{\rm \odot})$ is larger than $E_{\rm 2j}(100R_{\rm \odot})$ by more than a factor of two.  

At these radii, we find that the orbital energy of the unperturbed system is $E_{\rm orb}(20R_{\rm \odot}) = 8\times 10^{48} \erg$ and $E_{\rm orb}(100R_{\rm \odot})= 1.6\times 10^{48} \erg$. We see indeed that $E_{\rm 2j} \gg E_{\rm orb}$ in our simulations. Moreover, the binding energies of the unperturbed envelope down to a radius of $r=20R_{\rm \odot}$ and $r=100R_{\rm \odot}$ are $E_{\rm bind}(20R_{\rm \odot})=2.8\times 10^{49} \erg$ and $E_{\rm bind}(100R_{\rm \odot})=4.1\times 10^{48} \erg$, respectively. We note that in both cases $E_{\rm bind}(r)>E_{\rm orb}(r)$. Therefore, we can safely neglect the orbital energy in the simulations.  

\subsection{Neglecting dynamical effects}
\label{subsec:NeglectingDynamical}

We can neglect the spin-up of the envelope by the spiralling-in NS. The moment of inertia of the unperturbed spherical RSG envelope is $I_{\rm env,0}=5\times 10^{5} M_\odot R^2_\odot$. The moment of inertia of the NS on the surface of the RSG is larger, $I_{\rm orb,0} = M_{\rm NS} R^2_{\rm RSG} =9\times 10^{5} M_\odot R^2_\odot$.
However, after envelope expansion to a radius of $R_{\rm RSG} \simeq 4.1\times 10^3 - 9.6 \times 10^3 R_\odot$ depending on the value of $\zeta$,  we find a gross typical value of $I_{\rm env}\simeq 10^{8} M_\odot R^2_\odot \gg I_{\rm orb,0} $. We therefore expect, assuming a solid body rotation, the angular velocities of the envelope in regions where the NS spirals-in to be much lower than the Keplerian velocities at the respective orbital locations of the NS. 

We also neglect the dynamical response time of the envelope. Namely, we will use the stellar evolutionary model under the assumption of an hydrostatic equilibrium of the inflated RSG envelope. This is not an accurate assumption for the very outer zones of the inflated envelope. We inject the jets energy within 5 or 10 years. The dynamical time at the stellar radius $R_{\rm RSG}$ is $\tau_{\rm RSG,d} = (G \bar \rho)^{-1/2} = 5 (R_{\rm RSG}/10^4 R_\odot)^{3/2} \yr$ for an average density $\bar \rho$ of the RSG of mass $M_{\rm RSG} = 59.8 M_\odot$. In inner zones the dynamical times are much shorter and therefore the RSG has time to hydrostatically arrange itself during the spiraling-in phase. Namely, besides the very outer regions, the model is stable. As we are interested in the response of the envelope at radii of $r< 1000 R_\odot$ (see section \ref{sec:Results}), the effect of not including the dynamical response time of the envelope is not substantial. In any case, we expect that the outer parts of the inflated RSG envelope will be lost in a strong wind during the CEE. 

\subsection{Spherical symmetry and limited evolution time}
\label{subsec:SphericalSymmetry}
  
Three-dimensional hydrodynamical simulations have shown over more than three decades that CEE interaction is highly non-spherical even before we consider the effects of jets (e.g., \citealt{RasioLivio1996, LivioSoker1988, RickerTaam2008,  Passyetal2012, Nandezetal2014, Ohlmannetal2016, Iaconietal2017, MacLeodetal2018a, GlanzPerets2021}). The jets add another prominent non-spherical component to this asymmetry (e.g., \citealt {MorenoMendezetal2017, ShiberSoker2018, LopezCamaraetal2019, Shiberetal2019, LopezCamaraetal2020MN, Schreieretal2021}). With the numerical scheme we use, we are forced to assume a spherically symmetric energy deposition and envelope response. 
 
The non-spherical effects might be very significant at the final CEE phases (e.g., \citealt{Soker1992, Reichardtetal2019, GarciaSeguraetal2020, Zouetal2020}) and might include the formation of a circumbinary thick disk (e.g., \citealt{KashiSoker2011, ChenPodsiadlowski2017}). 
For that reason, and for neglecting the orbital energy (section \ref{subsec:NeglectingOrbital}) that increases to non-negligible values at small orbital separations, we do not continue the evolution to late phases when the orbital separation decreases below about $a=20 R_\odot$.

\section{Results: The effect of jets}
\label{sec:Results}
 
\begin{figure}
\begin{center}
\vspace*{-5.10cm}
\hspace*{-1.3cm}
\includegraphics[width=0.6\textwidth]{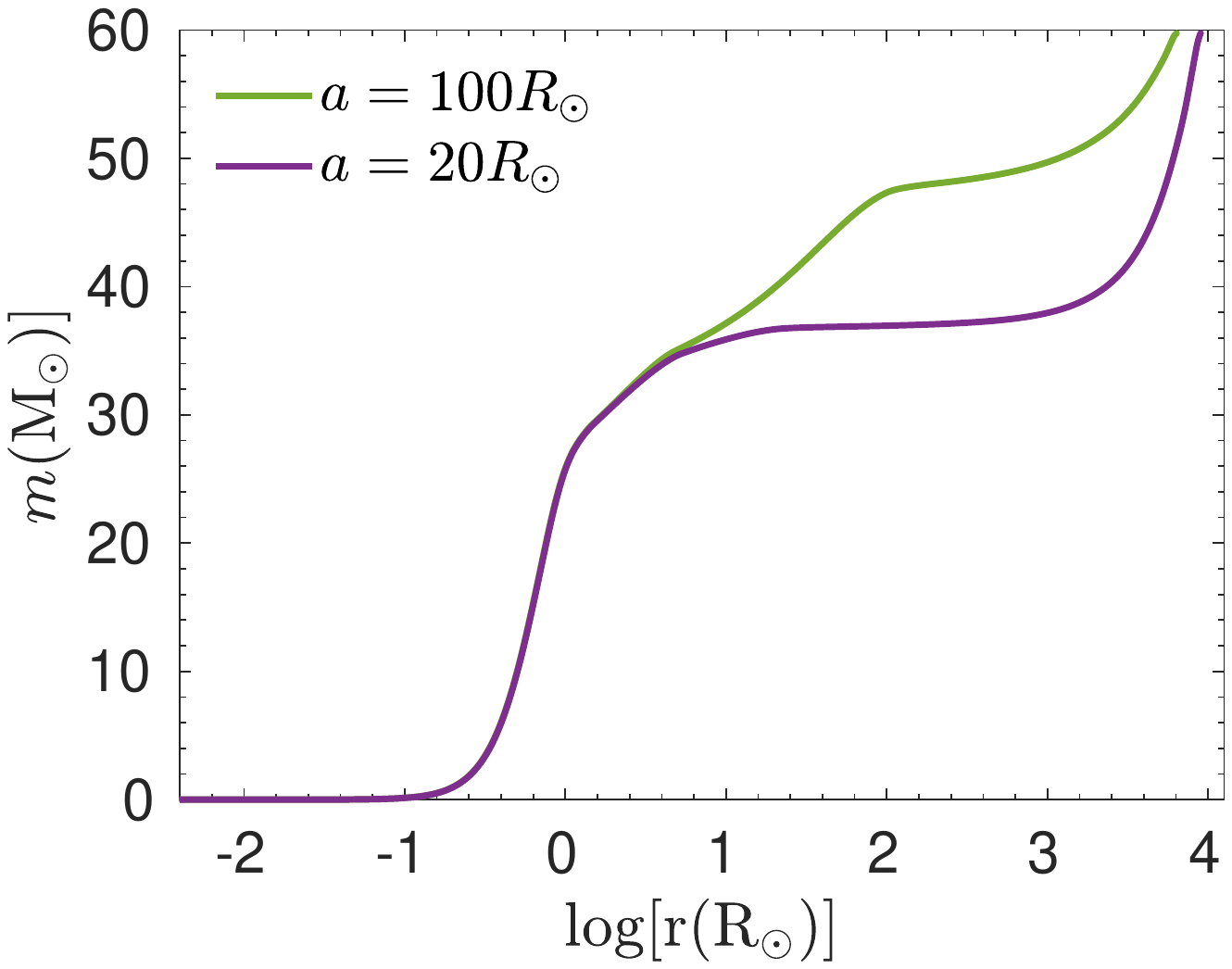}
\vspace*{-5.1cm}
\caption{Mass profile of the red supergiant (RSG) for $\zeta = 2.5 \times 10^{-3}$ in the first set of our simulations after $5 \yr$ when the orbital separation is $a=100R_{\rm \odot}$ (green curve), and in the second set of our simulations after $10 \yr$ when the orbital separation is $a=20R_{\rm \odot}$ (purple curve).
}
\label{fig:MassProfiles}
\end{center}
\end{figure}

\begin{figure}
\begin{center}
\vspace*{-5.0cm}
\hspace*{-1.3cm}
\includegraphics[width=0.6\textwidth]{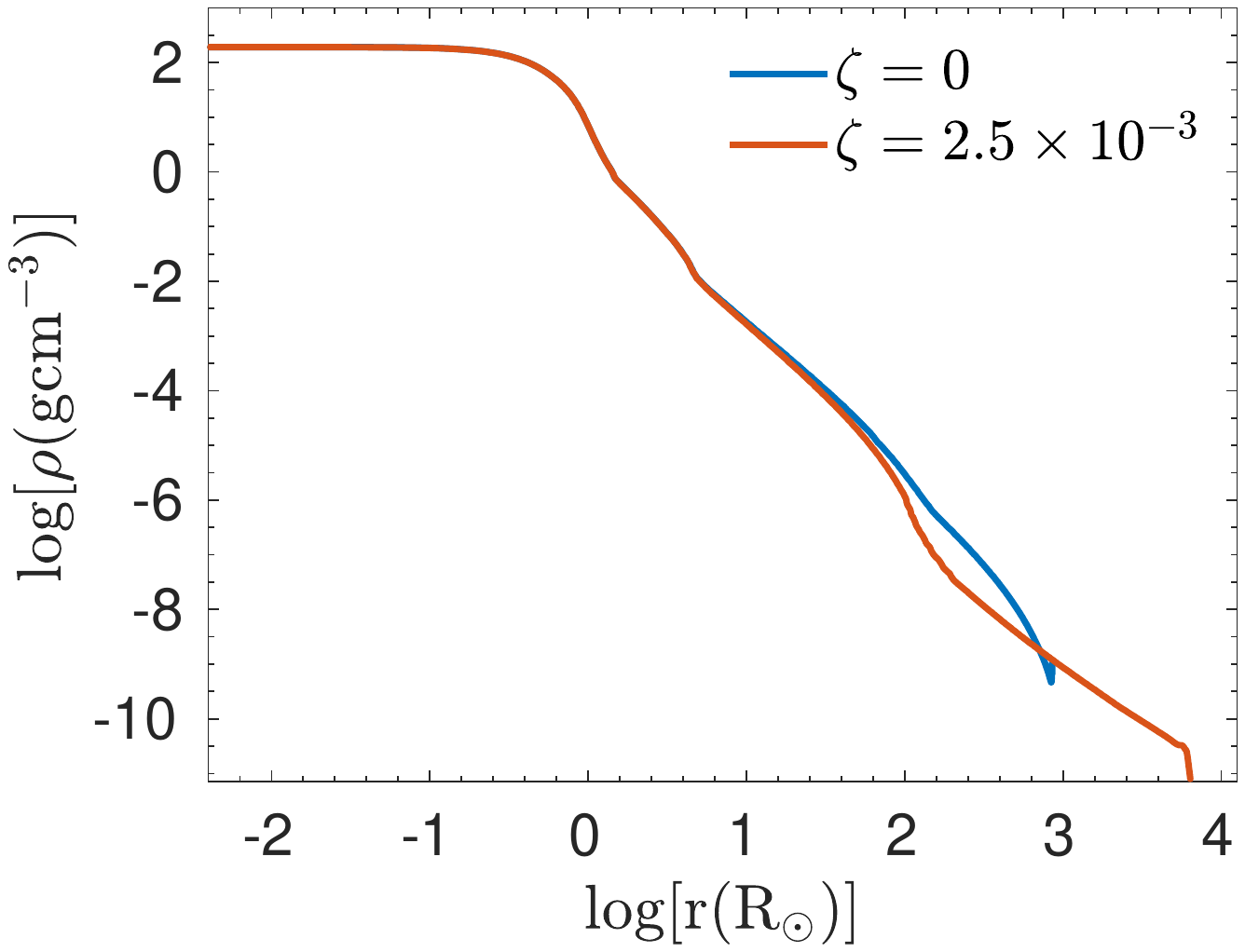}
\vspace*{-5.1cm}
\caption{Density profile of the red supergiant (RSG) in the first set of our simulations after $5 \yr$ when the orbital separation is $a=100R_{\rm \odot}$, for $\zeta = 2.5 \times 10^{-3}$ (red curve) and for $\zeta = 0$ (no energy injection; blue curve). 
}
\label{fig:Density100R}
\end{center}
\end{figure}

\begin{figure}
\begin{center}
\vspace*{-5.10cm}
\hspace*{-1.3cm}
\includegraphics[width=0.6\textwidth]{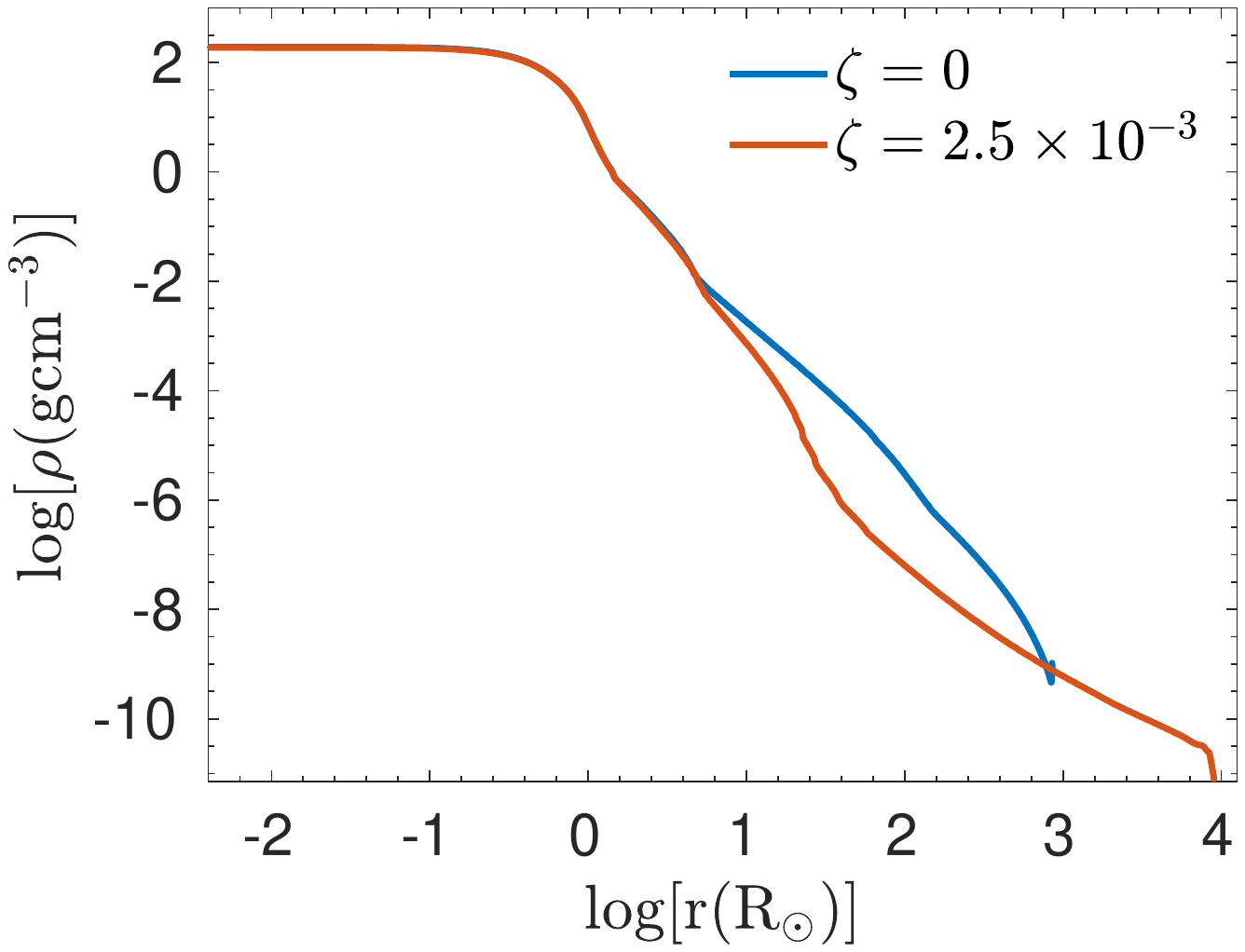}
\vspace*{-5.1cm}
\caption{Same as Fig. \ref{fig:Density100R} for our second set of simulations after $10 \yr$ of energy injection, when the orbital separation is $a=20R_{\rm \odot}$. 
}
\label{fig:Density20R}
\end{center}
\end{figure}

We determine the envelope structure at $t=\tau_{\rm j}$, i.e., at the end of jet-energy deposition in each one of our simulations. We present the results for $\zeta = 2.5 \times 10^{-3}$ for both $\tau_{\rm j} = 5 \yr$, where the NS spirals in down to $a = 100R_{\rm \odot}$ (Fig. \ref{fig:MassProfiles}, green curve; Fig. \ref{fig:Density100R}, red curve), and $\tau_{\rm j} = 10 \yr$, where the NS dives deeper to $a = 20R_{\rm \odot}$ (Fig. \ref{fig:MassProfiles}, purple curve; Fig. \ref{fig:Density20R}, red curve). For each case we also present the stellar density profile at these times for a case where we do not inject energy to the star (blue curve). 

Figs. \ref{fig:Density100R} and \ref{fig:Density20R} show the extended RSG envelope that the jets inflate. In the case where we mimic the spiraling-in of the NS down to $100R_{\rm \odot}$ (Fig. \ref{fig:Density100R}), the jets inflate the envelope of the giant star to a radius of $R_{\rm RSG} = 6.4\times 10^{3}R_{\rm \odot}$. When the NS spirals-in deeper, down to $20R_{\rm \odot}$ (Fig. \ref{fig:Density20R}), the envelope expands to an even higher radius of $R_{\rm RSG} = 9\times 10^{3}R_{\rm \odot}$.

Our aim is to determine the values of the feedback coefficient $\chi_{\rm j}$ under the assumption that the jet feedback mechanism operates in our scenario. In each of our simulations we inject energy to the envelope with a fixed value of $\zeta$ (equation \ref{eq:Zeta}). We use the densities depicted in Figs. \ref{fig:Density100R} and \ref{fig:Density20R} to find the values of $\chi_{\rm j}$ according to equation (\ref{eq:ChiJ}). Since there is a large uncertainty regarding the propagation of the jets inside the RSG, we compute $\chi_{\rm j}$ at three radii at the end of each simulation (5 or 10 years) according to equation (\ref{eq:ChiJ}). We limit the calculations to radii deep within the extended envelope, where dynamical effects can be neglected since the RSG envelope in these regions has time to rearrange itself in a hydrostatic equilibrium within a time much shorter than our simulation time (subsection \ref{subsec:NeglectingDynamical}). 

\begin{figure}
\begin{center}
\vspace*{-5.20cm}
\hspace*{-1.3cm}
\includegraphics[width=0.6\textwidth]{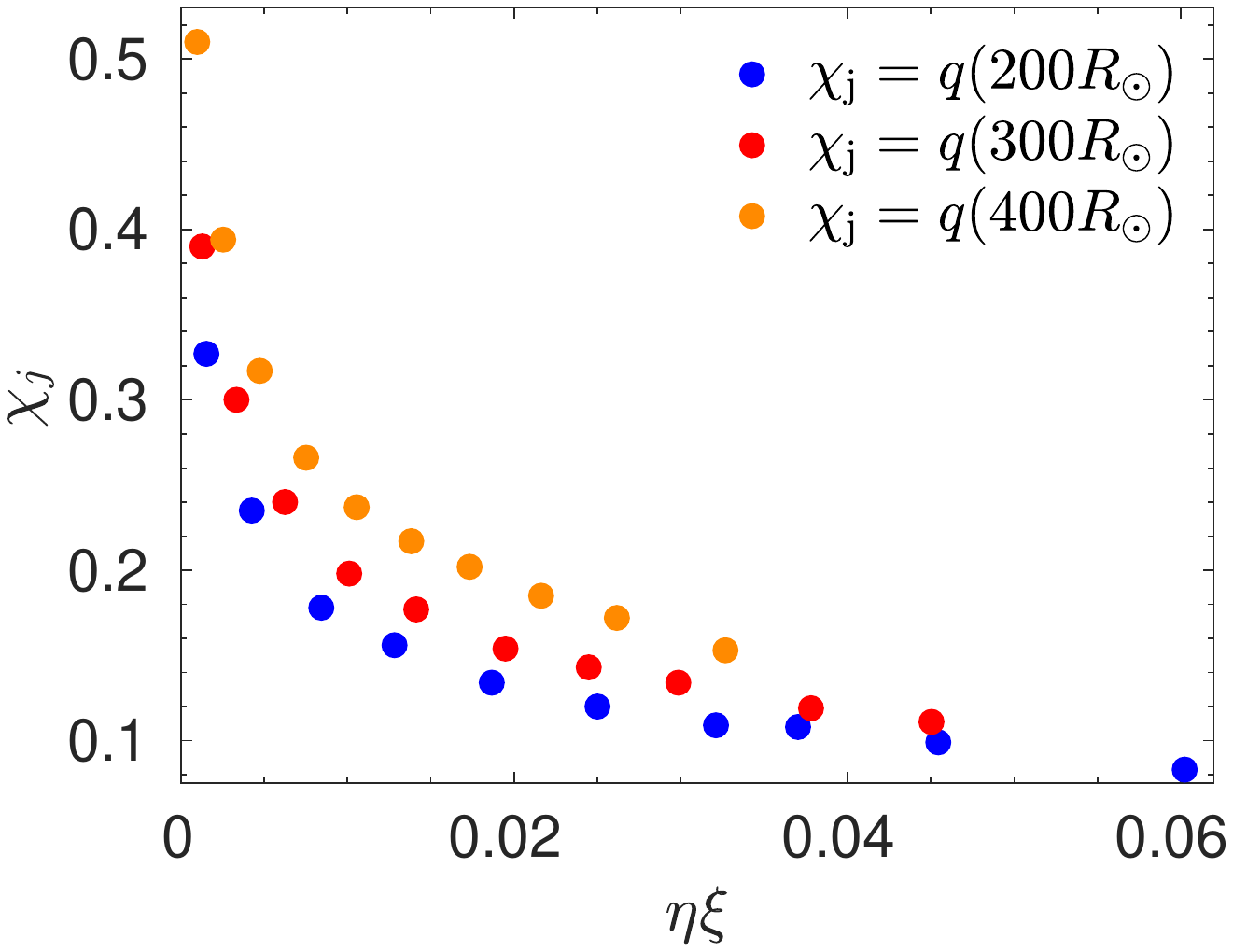}
\vspace*{-5.1cm}
\caption{The negative jet feedback coefficient $\chi_{\rm j}$ computed at $r=200R_{\rm \odot}$ (blue dots), at $r=300R_{\rm \odot}$ (red dots) and at $r=400R_{\rm \odot}$ (orange dots) for the first set of simulations ($a=100R_{\rm \odot}$, $\tau_{\rm j}=5\yr$). As $\zeta = \eta \chi_{\rm j} \xi$ each value of $\chi_{\rm j}$ corresponds to a different value of $\eta \xi$.  
}
\label{fig:Results100R}
\end{center}
\end{figure}

\begin{figure}
\begin{center}
\vspace*{-5.10cm}
\hspace*{-1.3cm}
\includegraphics[width=0.6\textwidth]{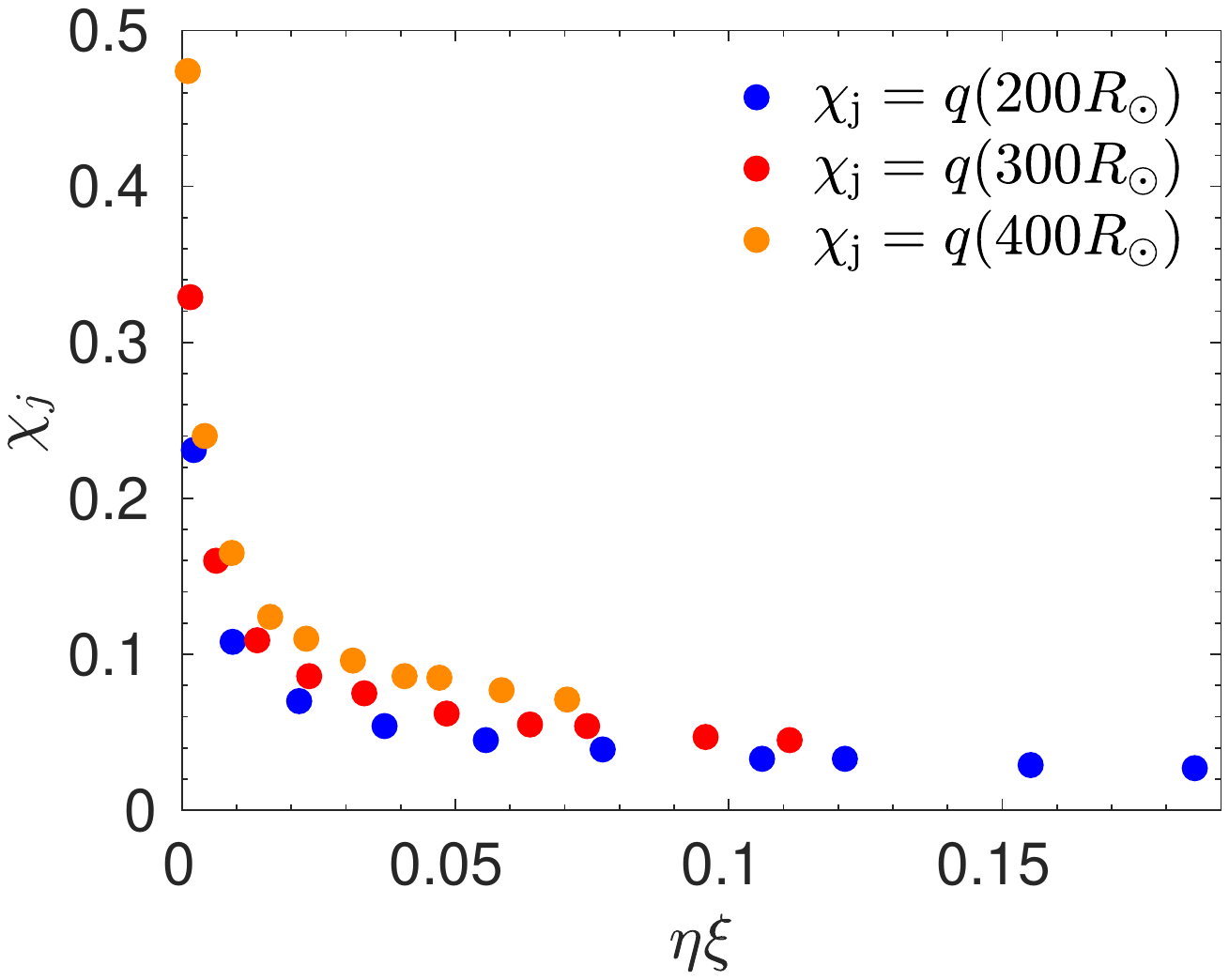}
\vspace*{-5.1cm}
\caption{Same as Fig. \ref{fig:Results100R} for our second set of simulations ($a=20R_{\rm \odot}$, $\tau_{\rm j}=10\yr$).
}
\label{fig:Results20R}
\end{center}
\end{figure}

In Figs. \ref{fig:Results100R} and \ref{fig:Results20R} we show the values of $\chi_{\rm j}$ computed at three different radii inside the RSG as we indicate in the inset. At these radii we calculate the density ratio of the perturbed to unperturbed envelope and find the values of $\chi_{\rm j}$ as function of $\eta \xi=\zeta/\chi_{\rm j}$. We find that the range of $\chi_{\rm j}$ that corresponds to the values of $\xi$ from previous hydrodynamical simulations is not much different when computing $\chi_{\rm j}$ in the three different radii above.

We note that although at the end of the first set of simulations the NS is at $a=100 R_\odot$ and at the end of the second set of simulations it is at $a=20 R_\odot$, we calculate the values of $\chi_{\rm j}$ (equation \ref{eq:ChiJ}) at somewhat larger radii of $r=200$, $300$ and $400 R_\odot$. The reason for this is that we inject the energy in spherical shells, and in our spherical stellar model we cannot take into account the local effect of the jets, which are larger near the NS location, including to zones inner to the NS orbit. Our spherical modeling is limited in its accuracy, and gives only the order of magnitude of the coefficient of the jet feedback mechanism $\chi_{\rm j}$.

As in most cases $\xi \simeq 0.1-0.5$ (e.g., \citealt{Livioetal1986}; \citealt{RickerTaam2008}; \citealt{Chamandyetal2018}), we can conclude from Figs. \ref{fig:Results100R} and \ref{fig:Results20R} that the effect of the jets in reducing the NS mass accretion rate in a CEE, i.e., the jets negative feedback mechanism coefficient that is compatible with the above values of $\xi$ is $\chi_{\rm j} \approx 0.04-0.3$. 

\section{Summary and discussion}
\label{sec:Summary}

We studied the effect of the negative jet feedback mechanism on the mass accretion rate by a NS that spirals-in inside the envelope of a RSG and launches jets, i.e., a CEJSN (or imposter) event. We performed one-dimensional simulations of a massive RSG star using the stellar evolution code \textsc{MESA} (section \ref{sec:Mimiking}). We mimicked the effect of jets that a NS launches inside the RSG envelope by depositing energy to the  envelope outside the NS location with a power according to equation (\ref{eq:JetsPower}). We performed two sets of simulations where we followed the NS spiraling-in down to orbital separations of either $a=100R_{\rm \odot}$ or $a=20R_{\rm \odot}$. For each set we performed simulations with different values $\zeta$ that is the ratio of jets power to the BHL accretion power from the unperturbed envelope (defined in equations \ref{eq:JetsPower} and \ref{eq:Zeta}). 

As we demonstrate for two cases with $\zeta=2.5 \times 10^{-3}$ in Figs. \ref{fig:Density100R} and \ref{fig:Density20R}, in all our simulations the jets substantially inflate the envelope. We take the negative jet feedback mechanism efficiency $\chi_{\rm j}$ to be the density ratio between the perturbed and unperturbed envelope (equation \ref{eq:ChiJ}). Due to the limitation of the spherical model that we used to mimic the highly non-spherical effects of the jets on the RSG envelope we take the density ratio somewhat outside the NS orbital radius at the end of the simulations. Since the exact radius is highly uncertain, we check the value of $\chi_{\rm j}$ at three radii as we indicate in Figs \ref{fig:Results100R} and \ref{fig:Results20R}.

Performing a a simple spherically symmetric mimicking of the energy of jets launched by a NS inside a RSG envelope in a CEJSN (or impostor) event, we could determine the reduction factor in the jets power due to the negative jet feedback mechanism. The main conclusion from this study is that under our assumptions, for values of $\eta \simeq 0.1$ as observations of jets indicate, and for $\xi \simeq 0.1-0.5$ as numerical simulations indicate, the jet feedback mechanism reduced the mass accretion rate onto a NS in the envelope of a RSG by a factor $\chi_{\rm j} \simeq 0.04-0.3$  (Figs. \ref{fig:Results100R} and \ref{fig:Results20R}).

\section*{Acknowledgments}
We thank an anonymous referee for very detailed comments that helped in improving our paper. This research was supported by a grant from the Israel Science Foundation (769/20). 

\textbf{Data availability}

The data underlying this article will be shared on reasonable request to the corresponding author. 


\begin{thebibliography}

\bibitem[Bondi \& Hoyle(1944)]{BondiHoyle1944} Bondi, H., \& Hoyle, F.\
1944, \mnras, 104, 273 

\bibitem[Broekgaarden \& Berger(2021)]{BroekgaardenBerger2021}  Broekgaarden, F.~S. \& Berger, E.\ 2021, arXiv:2108.05763 
\bibitem[Chamandy et al.(2018)]{Chamandyetal2018} Chamandy, L., Frank, A., Blackman, E.~G., et al.\ 2018, \mnras, 480, 1898. doi:10.1093/mnras/sty1950

\bibitem[Chen \& Podsiadlowski(2017)]{ChenPodsiadlowski2017} Chen, W.-C. \& Podsiadlowski, P.\ 2017, \apjl, 837, L19. doi:10.3847/2041-8213/aa624a

\bibitem[Chevalier(1993)]{Chevalier1993} Chevalier, R.~A.\ 1993, \apjl, 411, L33

\bibitem[Chevalier(2012)]{Chevalier2012} Chevalier, R.~A.\ 2012, \apjl, 752, L2

\bibitem[Fragos et al.(2019)]{Fragosetal2019} Fragos, T., Andrews, J.~J., Ramirez-Ruiz, E., Meynet, G., Kalogera, V., Taam, R.~E., \& Zezas, A., \ 2019, \apjl, 883, L45. doi:10.3847/2041-8213/ab40d1

\bibitem[Fryer \& Woosley(1998)]{FryerWoosley1998} Fryer, C.~L., \& Woosley, S.~E.\ 1998, \apjl, 502, L9

\bibitem[Garc{\'\i}a-Segura et al.(2020)]{GarciaSeguraetal2020} Garc{\'\i}a-Segura, G., Taam, R.~E., \& Ricker, P.~M.\ 2020, \apj, 893, 150. doi:10.3847/1538-4357/ab8006

\bibitem[Garc{\'\i}a et al.(2021)]{Garciaetal2021} Garc{\'\i}a, F., Simaz Bunzel, A., Chaty, S., Porter, E., \& Chassande-Mottin, E.\ 2021, arXiv:2103.03161

\bibitem[Gilkis et al.(2019)]{Gilkiselal2019} Gilkis, A., Soker, N., \& Kashi, A.\ 2019, \mnras, 482, 4233. doi:10.1093/mnras/sty3008

\bibitem[Glanz \& Perets(2021)]{GlanzPerets2021} Glanz, H. \& Perets, H.~B.\ 2021, \mnras, 500, 1921. doi:10.1093/mnras/staa3242

\bibitem[Grichener \& Soker(2019a)]{GrichenerSoker2019a} Grichener, A. \& Soker, N.\ 2019, \apj, 878, 24. doi:10.3847/1538-4357/ab1d5d

\bibitem[Grichener \& Soker(2019b)]{GrichenerSoker2019b} Grichener, A. \& Soker, N.\ 2019, arXiv:1909.06328

\bibitem[Grichener \& Soker(2021)]{GrichenerSoker2021} Grichener, A. \& Soker, N.\ 2021, arXiv:2101.05118

\bibitem[Hillel et al.(2021)]{Hilleletal2021} Hillel, S., Schreier, R., Shiber, S., \& Soker, N.\ 2021, in preparation

\bibitem[Hoang et al.(2020)]{Hoangetal2020} Hoang, B.-M., Naoz, S., \& Kremer, K.\ 2020, \apj, 903, 8. doi:10.3847/1538-4357/abb66a

\bibitem[Houck \& Chevalier(1991)]{HouckChevalier1991} Houck, J.~C., \& Chevalier, R.~A.\ 1991, \apj, 376, 234

\bibitem[Hoyle \& Lyttleton(1939)]{HoyleLyttleton1939} Hoyle, F., \& Lyttleton,
R.~A.\ 1939, Proceedings of the Cambridge Philosophical Society,
35, 405 

\bibitem[Iaconi et al.(2017)]{Iaconietal2017} Iaconi, R., Reichardt, T., Staff, J., De Marco, O., Passy, J.-C., Price, D., Wurster, J., \& Herwig, F.\ 2017, \mnras, 464, 4028

\bibitem[Ivanova et al.(2013)]{Ivanovaetal2013} Ivanova, N., Justham, S., Chen, X., et al.\ 2013, \aapr, 21, 59. doi:10.1007/s00159-013-0059-2

\bibitem[Kashi \& Soker(2011)]{KashiSoker2011} Kashi, A. \& Soker, N.\ 2011, \mnras, 417, 1466. doi:10.1111/j.1365-2966.2011.19361.x

\bibitem[Livio \& Soker(1988)]{LivioSoker1988} Livio, M., \& Soker, N.\ 1988, \apj, 329, 764

\bibitem[Livio et al.(1986)]{Livioetal1986} Livio, M., Soker, N., de Kool, M., \& Savonije, G.~J., \ 1986, \mnras, 222, 235. doi:10.1093/mnras/222.2.235

\bibitem[L{\'o}pez-C{\'a}mara et al.(2019)]{LopezCamaraetal2019} L{\'o}pez-C{\'a}mara, D., De Colle, F., \& Moreno M{\'e}ndez, E.\ 2019, \mnras, 482, 3646

\bibitem[L{\'o}pez-C{\'a}mara et al.(2020)]{LopezCamaraetal2020MN} L{\'o}pez-C{\'a}mara, D., Moreno M{\'e}ndez, E., \& De Colle, F.\ 2020, \mnras, 497, 2057

\bibitem[MacLeod \& Ramirez-Ruiz(2015a)]{MacLeodRamirezRuiz2015a} MacLeod, M., \&  {Ramirez-Ruiz}, E.\ 2015a, \apjl, 798, L19

\bibitem[MacLeod \& Ramirez-Ruiz(2015b)]{MacLeodRamirezRuiz2015b} MacLeod, M. \& Ramirez-Ruiz, E.\ 2015b, \apj, 803, 41. doi:10.1088/0004-637X/803/1/41

\bibitem[MacLeod et al.(2018)]{MacLeodetal2018a} MacLeod, M., Ostriker, E.~C., \& Stone, J.~M.\ 2018, \apj, 863, 5. doi:10.3847/1538-4357/aacf08

\bibitem[Mapelli(2020)]{Mapelli2020} Mapelli, M.\ 2020, Frontiers in Astronomy and Space Sciences, 7, 38. doi:10.3389/fspas.2020.00038

\bibitem[Moreno M{\'e}ndez et al.(2017)]{MorenoMendezetal2017} Moreno M{\'e}ndez, E., L{\'o}pez-C{\'a}mara, D., \& De Colle, F.\ 2017, \mnras, 470, 2929 

\bibitem[Nandez et al.(2014)]{Nandezetal2014} Nandez, J.~L.~A., Ivanova, N., \& Lombardi, J.~C., Jr.\ 2014, \apj, 786, 39

\bibitem[Ohlmann et al.(2016)]{Ohlmannetal2016} Ohlmann, S.~T., R{\"o}pke, F.~K., Pakmor, R., \& Springel, V.\ 2016, \apjl, 816, L9

\bibitem[Passy et al.(2012)]{Passyetal2012} Passy, J.-C., De Marco, O., Fryer, C.~L., et al.\ 2012, \apj, 744, 52. doi:10.1088/0004-637X/744/1/52

\bibitem[Paxton et al.(2010)]{Paxtonetal2010} Paxton, B., Bildsten, L., Dotter, A., et al.\ 2010, Astrophysics Source Code Library. ascl:1010.083

\bibitem[Paxton et al.(2013)]{Paxtonetal2013} Paxton, B., Cantiello, M., Arras, P., et al.\ 2013, \apjs, 208, 4 

\bibitem[Paxton et al.(2015)]{Paxtonetal2015} Paxton, B., Marchant, P., Schwab, J., et al.\ 2015, \apjs, 220, 15 

\bibitem[Paxton et al.(2018)]{Paxtonetal2018} Paxton, B., Schwab, J., Bauer, E.~B., et al.\ 2018, The Astrophysical Journal Supplement Series, 234, 34.

\bibitem[Paxton et al.(2019)]{Paxtonetal2019} Paxton, B., Smolec, R., Schwab, J., et al.\ 2019, \apjs, 243, 10, 

\bibitem[Rasio \& Livio(1996)]{RasioLivio1996} Rasio, F.~A., \& Livio, M.\ 1996, \apj, 471, 366

\bibitem[Reichardt et al.(2019)]{Reichardtetal2019} Reichardt T.~A., De Marco O., Iaconi R., Tout C.~A., Price D.~J., 2019, MNRAS, 484, 631

\bibitem[Ricker \& Taam(2008)]{RickerTaam2008} Ricker, P.~M., \& Taam, R.~E.\ 2008, \apjl, 672, L41

\bibitem[Ricker \& Taam(2012)]{RickerTaam2012} Ricker, P.~M. \& Taam, R.~E.\ 2012, \apj, 746, 74. doi:10.1088/0004-637X/746/1/74

\bibitem[Sand et al.(2020)]{Sandetal2020} Sand, C., Ohlmann, S.~T., Schneider, F.~R.~N., Pakmor, R., \& R{\"o}pke, F.~K., \ 2020, \aap, 644, A60. doi:10.1051/0004-6361/202038992

\bibitem[Schreier et al.(2021)]{Schreieretal2021} Schreier, R., Hillel, S., Shiber, S., \& Soker, N.\ 2021, 

\bibitem[Schr{\o}der et al.(2020)]{Schroderetal2020} Schr{\o}der, S.~L., MacLeod, M., Loeb, A., et al.\ 2020, \apj, 892, 13

\bibitem[Shiber et al.(2019)]{Shiberetal2019} Shiber, S., Iaconi, R., De Marco, O., \& Soker, N.\ 2019, \mnras, 488, 5615. doi:10.1093/mnras/stz2013

\bibitem[Shiber et al.(2016)]{Shiberetal2016} Shiber, S., Schreier, R., \& Soker, N.\ 2016, RAA, 16, 117

\bibitem[Shiber \& Soker(2018)]{ShiberSoker2018} Shiber, S. \& Soker, N.\ 2018, \mnras, 477, 2584. doi:10.1093/mnras/sty843

\bibitem[Soker(1992)]{Soker1992} Soker, N.\ 1992, \apj, 386, 190

\bibitem[Soker(2016)]{Soker2016Rev} Soker, N.\ 2016, \nar, 75, 1. doi:10.1016/j.newar.2016.08.002

\bibitem[Soker(2021)]{Soker2021} Soker, N.\ 2021, \mnras, 504, 5967. doi:10.1093/mnras/stab1275

\bibitem[Soker \& Gilkis(2018)]{SokerGilkis2018} Soker, N. \& Gilkis, A.\ 2018, \mnras, 475, 1198. doi:10.1093/mnras/stx3287

\bibitem[Soker et al.(2019)]{Sokeretal2019} Soker, N., Grichener, A., \& Gilkis, A.\ 2019, \mnras, 484, 4972. doi:10.1093/mnras/stz364

\bibitem[Vigna-G{\'o}mez et al.(2018)]{VignaGomezeal2018} Vigna-G{\'o}mez, A., Neijssel, C.~J., Stevenson, S., et al.\ 2018, \mnras,

\bibitem[Zevin et al.(2021)]{Zevinetal2021} Zevin, M., BavZevin, M., Bavera, S.~S., Berry, C.~P.~L., et al.\ 2021, \apj, 910, 152

\bibitem[Zou et al.(2020)]{Zouetal2020} Zou, Y., Frank, A., Chen, Z., et al.\ 2020, \mnras, 497, 2855. doi:10.1093/mnras/staa2145
 
\end{thebibliography}
\end{document}